# Expanding RCWA capabilities with advanced S-matrix algorithms

P. Lalanne[1*] and J.-P. Hugonin[2]

[1]Laboratoire Photonique Numérique et Nanosciences (LP2N), Université de Bordeaux, Institut d'Optique Graduate School, CNRS, Talence, France

[2]Université Paris-Saclay, Institut d'Optique Graduate School, CNRS, Laboratoire Charles Fabry, 91127 Palaiseau, France

[*]Corresponding author: philippe.lalanne@institutoptique.fr

## Abstract

Modal methods are particularly well suited to layered photonic structures because electromagnetic propagation within each layer is described analytically. Here, we introduce a simple reformulation of the scattering-matrix (S-matrix) approach in which the S-matrix is defined as an intrinsic property of an individual layer, independently of its neighboring interfaces. This separation between layer propagation and interface coupling preserves the numerical stability of the conventional S-matrix formalism while providing a more modular description of multilayer structures. It also enables direct computation of the scattering coefficients between external plane waves and Bloch modes, as well as between Bloch modes themselves, providing generalized Fresnel coefficients for periodic interfaces. In addition, the reformulation allows the optical response to be evaluated simultaneously for many layer thicknesses or incident wavevectors, with only a modest computational overhead compared with a single calculation. Implemented in the RETICOLO freeware, these capabilities provide a practical framework for modal analysis, parameter sweeps, and the design of layered photonic structures.

## 1. Introduction

Layered structures form the backbone of many photonic devices, including dielectric and metallic mirrors, thin-film coatings, diffraction gratings, and integrated optical components. Their optical response arises from multiple scattering and interference at successive interfaces, making its accurate prediction a central problem in computational electromagnetics. Beyond prediction, modern design also requires numerical methods that provide physical insight and enable rapid optimization.

Among the numerical approaches developed over the past decades, the scattering-matrix (S-matrix) formalism, which originated in mode-matching techniques in microwave engineering [1], has become one of the most versatile and robust frameworks for analyzing layered structures [2-5]. Unlike transfer-matrix formulations, which may suffer from numerical instabilities caused by exponentially growing evanescent waves, the S-matrix propagates only incoming and outgoing wave amplitudes. It is therefore intrinsically stable, naturally accommodates arbitrary multilayer structures, and provides direct access to reflection, transmission, absorption, and diffraction efficiencies.

The S-matrix formalism is largely independent of the numerical discretization adopted within each layer. It can be combined with modal methods such as the rigorous coupled-wave analysis (RCWA) [2-5], finite-difference and finite-element modal methods [6-7], or classical mode matching [1]. In all cases, Maxwell's equations are first solved independently within each layer to determine the propagating and evanescent eigenmodes, after which the global response is obtained by recursively assembling the layer S-matrices.

Despite its maturity, we believe that the computational potential of the S-matrix formalism has not yet been fully exploited. For instance, the analytical dependence of the S-matrix on layer thickness offers opportunities to accelerate parameter sweeps, model partially coherent propagation, and develop more efficient design strategies. The purpose of this work is to show how a simple reformulation of the S-matrix algorithm can substantially extend these capabilities.

We illustrate these developments using the RCWA implemented in the freely available RETICOLO software [8], which relies on the eigensolver algorithms described in Refs. [9,10]. The present work extends the public version of the software, offering capabilities that were previously available only in an in-house version [11]. Although all examples are presented within the RCWA framework, the proposed formulation is not specific to RCWA and can be applied to any modal method based on the propagation and coupling of eigenmodes in layered media.

In conventional RCWA implementations, the S-matrix is usually expressed in the modal basis and associated with a layer together with its adjacent interfaces [4]. Consequently, its definition depends on the neighboring layers through the modal bases used on either side of the interface. In contrast, we express the S-matrix directly in the Fourier basis and associate it exclusively with propagation through an individual layer. The S-matrix thus becomes an *intrinsic property of the layer*, relating the tangential electromagnetic fields at its two boundaries independently of the nature of the adjacent layers.

Section 2 introduces the proposed S-matrix reformulation and emphasizes the separation between intrinsic layer propagation and interface-dependent modal coupling. As demonstrated throughout the paper, the reformulation preserves the numerical stability of the classical S-matrix algorithm and its recursive assembly through the Redheffer star product [12], while providing a more modular description of layered structures that enables several new computational strategies.

Section 3 exploits the reformulated S-matrix to compute the coupling coefficients between external radiation channels and Bloch modes, as well as between Bloch modes themselves. This rigorous description of energy exchange at periodic interfaces extends RCWA from a predictive solver of global optical quantities to a modal analysis framework, in which the response of layered systems can be interpreted in terms of effective indices, interfacial boundary conditions, Bloch-mode excitation, and mode conversion.

Section 4 exploits the analytical dependence of the S-matrix on layer thickness and presents a vectorized implementation that simultaneously computes the S-matrices associated with many thicknesses of a given layer with only a modest overhead compared with a single-thickness calculation. This approach considerably accelerates parameter sweeps and optimization procedures and provides a simple practical treatment of partially coherent propagation in thick layers through thickness averaging.

Conventional RCWA implementations, such as those of S4 [13] and RETICOLO [8], compute the diffraction efficiencies associated with a single incident wave. In Section 5, we show that all scattering coefficients between every incident and diffracted channel can instead be evaluated simultaneously. This capability is particularly attractive for analyzing structures with large unit cells, such as disordered metasurfaces modeled through artificial periodicity [14].

Finally, Section 6 concludes the paper.

Annex 1 summarizes the new RETICOLO functionalities introduced in this work and provides practical guidelines for their use. The main scripts dedicated to S-matrix manipulations are described, and the MATLAB scripts used to generate all figures are available from the authors upon request. These documented scripts are intended both to reproduce the results presented here and to serve as practical examples for users developing their own calculations.

## 2. Reformulation of the S-Matrix algorithm

We consider a structure that is periodic along the $x$- and $y$-directions with periods $a_x$ and $a_y$, respectively. The permittivity and permeability are allowed to vary with all three spatial coordinates. Owing to the periodicity in the transverse directions, the propagation axis $z$ plays a particular role, and Maxwell's equations can be written as a first-order differential equation

$$\frac{d}{dz}\begin{bmatrix} E_x \\ E_y \\ H_x \\ H_y \end{bmatrix} = \mathbf{H}(x,y,z)\begin{bmatrix} E_x \\ E_y \\ H_x \\ H_y \end{bmatrix}, \quad (1)$$

where $\mathbf{H}$ is a differential operator acting only on the transverse coordinates $x$ and $y$, while depending parametrically on $z$. Solving Maxwell's equations therefore amounts to determining the state vector $\mathbf{X}(z) = [E_x, E_y, H_x, H_y]$ whose dimension may itself depend on $z$ in the most general formulations [7,15]. Once $\mathbf{X}(z)$ is known, the electromagnetic field is completely determined throughout the structure.

In this work, we restrict ourselves to layered structures, for which the operator $\mathbf{H}$ is piecewise constant along the propagation direction. In other words, each layer is homogeneous with respect to $z$, although it may exhibit an arbitrary periodic pattern in the transverse plane. The same formalism could readily be extended to structures that are only piecewise periodic along $z$, for example stacks containing an integer number of longitudinal periods [16], but this additional generality is not required here.

Let us denote by $z_p, p = 1,2 \ldots$ the layer interfaces. For each layer, Eq. (1) can be solved in either a numerical basis or a modal basis. In the RCWA, for example, the numerical basis is a truncated Fourier basis, whereas the modal basis consists of plane waves in homogeneous layers and Bloch modes in laterally periodic layers. The same framework also applies to supercell approaches employing artificial periodicity, such as those used for modeling integrated photonic devices [16,17].

### 2.1 Classical S-Matrix formalism

The scattering matrix (S-matrix) is a powerful formalism for describing electromagnetic wave propagation through multilayered and periodic structures. Classically [4], the S-matrix connects the amplitudes of incoming and outgoing modes (either Bloch modes or plane waves) at the upper and lower bounding interfaces, $z = z_p^+$ and $z = z_{p+1}^+$, of each layer $p$, encompassing propagation through a single layer and a single interface,

$$\begin{bmatrix} \mathbf{b}^- \\ \mathbf{a}^+ \end{bmatrix} = \begin{bmatrix} \mathrm{S}_{11} & \mathrm{S}_{12} \\ \mathrm{S}_{21} & \mathrm{S}_{22} \end{bmatrix}\begin{bmatrix} \mathbf{a}^- \\ \mathbf{b}^+ \end{bmatrix}, \quad (2)$$

where $\mathbf{a}^-$ and $\mathbf{b}^+$ denote the incident modal amplitudes from the upper and lower sides, respectively, while $\mathbf{b}^-$ and $\mathbf{a}^+$ represent the corresponding reflected and transmitted amplitudes (Fig. 1).

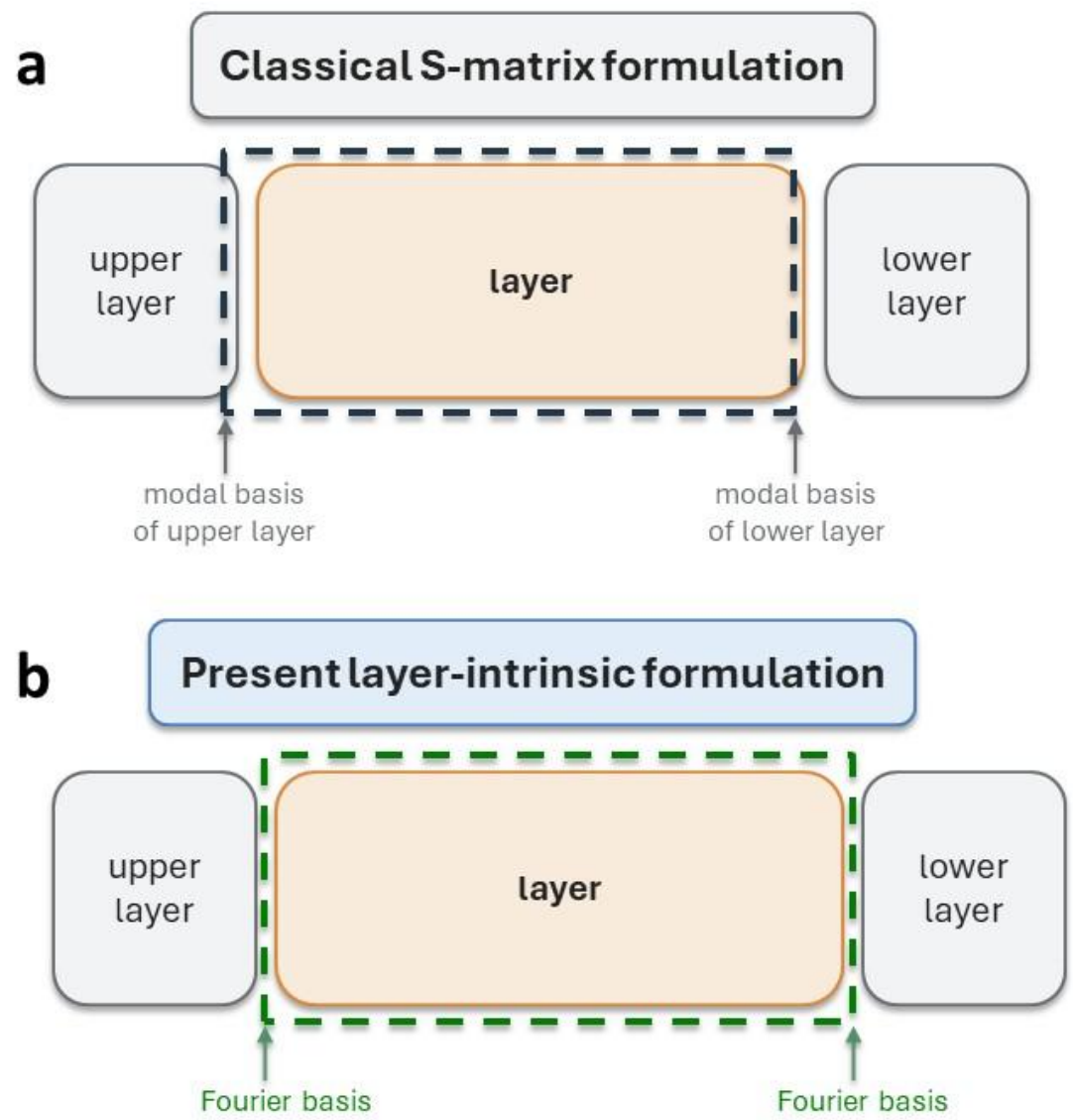

**Figure 1. Classical and layer-intrinsic S-matrix formulations. a** The classical S-matrix contains interface coupling and layer propagation and is expressed in the modal bases of the neighboring layers. **b** In the present formulation, the S-matrix is defined in the Fourier basis and describes propagation through the layer alone, independently of the neighboring layers. Interface-dependent modal coupling is introduced separately through basis transformations, and individual layer S-matrices can subsequently be assembled using the Redheffer star product.

Compared with transfer-matrix approaches, the S-matrix formulation exhibits superior numerical stability, particularly for thick structures and highly evanescent modes. This property makes it well suited for rigorous electromagnetic simulations based on modal methods, such as the RCWA considered in this work.

The S-matrix formalism constitutes the core framework of RCWA codes for describing propagation, reflection, and transmission within periodic structures. Once the modal basis of each layer has been determined through eigenvalue diagonalization, the corresponding layer S-matrix is computed and subsequently combined with those of adjacent layers using the Redheffer star product, allowing complex multilayer structures to be assembled from elementary building blocks while preserving numerical robustness [4].

## 2.2 Present S-Matrix formalism

The classical formulation is well suited to conventional multilayer gratings and remains the standard approach in most RCWA implementations. However, a different viewpoint becomes advantageous for more complex structures comprising many layers, such as tapered photonic-crystal waveguides and other integrated photonic devices.

Instead of expressing the S-matrix in the modal basis and associating it with a layer together with its adjacent interfaces, we express it directly in the Fourier basis and associate it solely with propagation through an individual layer. The resulting S-matrix relates the tangential electromagnetic fields at $z = z_p$ and $z = z_{p+1}$, independently of the neighboring layers. It therefore becomes an intrinsic property of the layer rather than of its interfaces.

This distinction is important. In the modal formulation, the S-matrix depends on the modal bases chosen on both sides of the layer and therefore implicitly depends on neighbouring layers. In contrast, the Fourier-basis formulation defines an intrinsic scattering operator for the layer itself. A given layer can therefore be characterized once and reused in arbitrary multilayer assemblies without recomputing interface-dependent quantities.

This conceptual modification, previously available only in the in-house version of RETICOLO [8], has proved particularly useful for the study of propagation and emission in photonic-crystal waveguides and cavities [18,19], where numerous layers with different neighbours must be combined repeatedly. In such situations, the Fourier-basis formulation considerably reduces the number of modal conversions and S-matrix products required during the computation.

Because the tangential electromagnetic fields are continuous across interfaces, one has

$$\mathbf{X}(z_p^+) = \mathbf{X}(z_p^-) \equiv \mathbf{X}(z_p), \tag{3}$$

so that no distinction between the "+" and "−" sides of an interface is required when working in the Fourier basis.

To simplify, let us introduce the notations $\mathbf{E} = \begin{bmatrix} E_x \\ E_y \end{bmatrix}$ and $\mathbf{H} = \begin{bmatrix} H_x \\ H_y \end{bmatrix}$. Starting from the standard T matrix, which relates the tangential fields between $z = z_p$ and $z = z_{p+1}$ by

$$\begin{bmatrix} \mathbf{E}(z_{p+1}) \\ \mathbf{H}(z_{p+1}) \end{bmatrix} = \begin{bmatrix} \mathrm{T}_{11} & \mathrm{T}_{12} \\ \mathrm{T}_{21} & \mathrm{T}_{22} \end{bmatrix} \begin{bmatrix} \mathbf{E}(z_p) \\ \mathbf{H}(z_p) \end{bmatrix}, \tag{4}$$

we define the S-matrix by

$$\begin{bmatrix} \mathbf{E}(z_{p+1}) \\ \mathbf{H}(z_p) \end{bmatrix} = \mathrm{S} \begin{bmatrix} \mathbf{E}(z_p) \\ \mathbf{H}(z_{p+1}) \end{bmatrix}. \tag{5}$$

We find

$$\mathrm{S} = \begin{bmatrix} \mathrm{I} & -\mathrm{T}_{12} \\ 0 & -\mathrm{T}_{22} \end{bmatrix}^{-1} \begin{bmatrix} \mathrm{T}_{11} & 0 \\ \mathrm{T}_{21} & -\mathrm{I} \end{bmatrix}. \tag{6}$$

In practice, Eq. (6) avoids the explicit computation of $\mathrm{T}_{22}^{-1}$ and, as we observed in several numerical experiments, provides better numerical stability than the alternative classical expression $\mathrm{S} = \begin{bmatrix} \mathrm{T}_{11} - \mathrm{T}_{12}\mathrm{T}_{22}^{-1}\mathrm{T}_{21} & \mathrm{T}_{12}\mathrm{T}_{22}^{-1} \\ -\mathrm{T}_{22}^{-1}\mathrm{T}_{21} & \mathrm{T}_{22}^{-1} \end{bmatrix}$. The propagation through a stack of layers is then performed with S-matrix products, using the Redheffer star product.

The advantage of this reformulation is not merely algebraic. In the conventional formulation, changing one of the media adjacent to a layer changes the modal basis and hence the associated scattering matrix. In the present formulation, the operator describing propagation through the layer is unchanged. Interfaces are treated separately through basis transformations and S-matrix composition. A layer can therefore be computed once and reused in arbitrary environments. This separation between layer propagation and interface coupling is the key to all the developments presented below.

### 2.3 Identification of incoming and outgoing modes

All these matrices are defined in the Fourier basis. When one wishes to compute scattering coefficients between a plane wave in a homogeneous medium and a Bloch mode of a structured layer, or between two Bloch modes, the S-matrix must be transformed into the corresponding modal basis.

The conversion between the Fourier basis and a modal basis is performed straightforwardly using a change-of-basis matrix [20,21]. It however requires the identification of incoming and outgoing modes. For propagating modes, the classification can be based on the direction of energy transport. More precisely, the sign of the $z$-component of the Poynting vector determines whether the mode is incoming or outgoing. For evanescent modes at real frequencies, outgoing modes are conventionally defined as those whose amplitude decreases exponentially away from the interface.

More complicated situations arise when the frequency is complex. Such cases occur naturally in the computation of quasinormal modes (QNMs) [22], but also in scattering calculations performed directly at complex frequencies [23-25]. In this context, even evanescent waves may become exponentially increasing along their direction of propagation. The distinction between incoming and outgoing modes therefore requires a rigorous sign convention and a consistent definition of the square root appearing in the propagation constants. Throughout this work, the square root is defined according to the convention

$$Re(\sqrt{x}) + Im(\sqrt{x}) > 0, \qquad (7)$$

as justified in Section 18.4.3 of Ref. [26].

## 3. Scattering coefficients between plane waves and Bloch modes

For homogeneous media, Fresnel coefficients quantify the coupling between incident and reflected or transmitted plane waves. They explain impedance matching, reflection suppression, and Fabry–Perot resonances. Periodic media possess analogous quantities, namely the scattering coefficients that govern the coupling between external plane waves and Bloch modes, as well as between Bloch modes themselves. These generalized Fresnel coefficients reveal the internal energy pathways of structured devices, identifying which Bloch modes are excited, how they exchange energy at interfaces, and which interference mechanisms are responsible for the observed optical response.

This physical interpretation, in which a periodic layer is described by dispersive Bloch modes together with generalized interfacial boundary conditions, goes beyond the mean-field description of homogeneous effective media. It underlies the coupled-Bloch-mode models developed in Refs. [27-36], where broadband reflection, polarization selectivity, and sharp spectral anomalies were explained in terms of the excitation, propagation, and repeated coupling of only a few Bloch modes, rather than through global reflection and transmission spectra alone. These studies demonstrated that RCWA contains considerably more physical information than is usually extracted in conventional workflows.

In this section, we show how the reformulated S-matrix provides direct access to these modal scattering coefficients. Through a simple numerical example, we demonstrate how they can be used to visualize and quantify the transfer of electromagnetic energy between external plane waves and Bloch modes, thereby transforming RCWA from a predictive solver into a practical tool for modal engineering.

### 3.1 Dominant Bloch modes

We illustrate the proposed approach using the design of a nanopatterned silicon layer providing broadband antireflection. The structure consists of a periodic array of silicon nanowires deposited on a semi-infinite silicon substrate. The grating period is 130 nm, the grating thickness is 120 nm, and the nanowire width is 80 nm. The silicon dispersion is taken from Ref. [37].

The computed spectrum (Fig. 2a) features a pronounced minimum near $\lambda = 440$ nm together with low reflectance throughout the infrared, in agreement with recent experimental observations on related two-dimensional silicon gratings [38]. This antireflection mechanism is commonly interpreted in terms of Mie resonances supported by the individual nanowires [38,39-41] or by their mutual interaction within the array [39,42-43].

The isolated-particle picture has proved remarkably successful for understanding many properties of dielectric nanoresonators [44]. In a periodic array, however, the resonators no longer behave independently. Their response results from both near-field interactions and long-range radiative coupling mediated by the periodic lattice [45], giving rise to collective resonances often referred to as lattice Mie resonances [39]. Although this interpretation is physically meaningful and can be modeled rigorously [46], it does not naturally reveal the optical pathways through which energy is transferred inside the structure.

Here we adopt a complementary viewpoint based on Bloch modes. Rather than describing the response in terms of resonances of individual meta-atoms, we interpret it as the excitation, propagation, and mutual coupling of Bloch modes at the interfaces between the grating, the substrate, and the superstrate. This description follows directly from the modal scattering coefficients introduced above and provides a rigorous yet intuitive picture of the internal energy flow responsible for the observed antireflection.

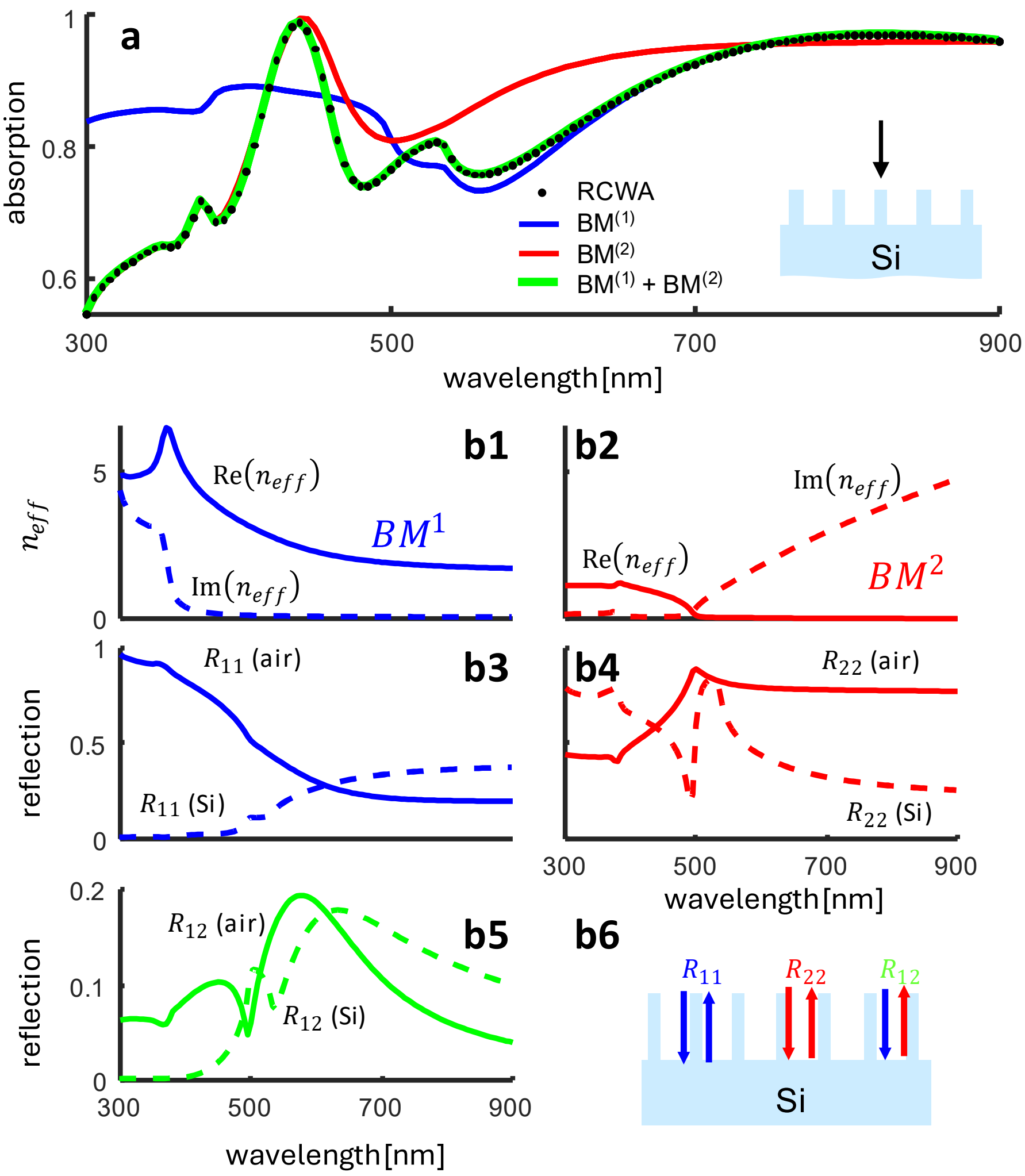

**Figure 2. Coupled Bloch-mode analysis of broadband antireflection in a silicon grating. a** Absorption spectrum under TM polarization at normal incidence. Inset: periodic array of silicon nanowires on a semi-infinite silicon substrate. **b1-b2** Effective index of Bloch modes $BM^1$ and $BM^2$, respectively. **b3-b5** Generalized reflection Fresnel coefficients at the air and Si interfaces for $BM^1$ and $BM^2$. **b6** Sketch of the reflection coefficients (blue color holds for $BM^1$ and red color for $BM^2$). The figure can be reproduced with the documented MATLAB script **fig2broadband_absorption.m**, available upon request. In addition to the reflected intensities, the script provides the reflected amplitudes (generalized Fresnel coefficients). See also the companion tutorial script **fig2broadband_absorption.m**.

We begin by computing the propagating Bloch modes of the grating. In the spectral range of interest, only two modes, denoted $BM^{(1)}$ and $BM^{(2)}$, a propagate. Their normalized propagation constants, $n_{eff} = ck_z/\omega$, i.e. their effective indices, computed with RETICOLO are plotted in Figs. 2(b1) and 2(b2). The fundamental mode, $BM^{(1)}$, exhibits a large effective index and can be viewed as the supermode associated with the fundamental guided mode of a thin silicon slab surrounded by air. As the wavelength decreases, the increasing confinement of the mode results in a larger effective index.

The second mode, $BM^{(2)}$, exhibits a markedly different behavior. It is evanescent for wavelengths above 500 nm but becomes propagative at shorter wavelengths, with an effective index of approximately 0.7 at $\lambda = 450$ nm. Such an unusually low effective index (smaller than that of air) is readily understood from a ray-optics picture [29]. The mode propagates along a zigzag trajectory, with steep propagation angles in the air regions and nearly horizontal propagation inside the high-index silicon, resulting in a small longitudinal propagation constant. It may lead to a grating-waveguide resonance with some leakage.

### 3.2 Bloch mode scattering coefficients

To interpret the optical response of the grating, we first adopt the simplest possible picture and assume that only a single Bloch mode propagates inside the periodic layer. The grating is then treated as an effective thin film supporting one propagating mode that undergoes multiple reflections at the air/grating and grating/silicon interfaces. Its reflectance is obtained from an Airy-like formula (Section 15.4 of Ref. [26]), where the dispersive propagation phase is determined by the effective index of the selected Bloch mode and the Fresnel coefficients are replaced by the modal scattering coefficients.

These generalized Fresnel coefficients are computed directly from the reformulated S-matrix approach. The reflection coefficients of $BM^{(1)}$ and $BM^{(2)}$ at the air and silicon interfaces are shown in Figs. 2(b3) and 2(b4). In practice, the single-mode approximation is implemented by truncating the modal basis to the selected Bloch mode (see Annex 1 and the script **fig2broadband_absorption.m**).

The resulting reflectance spectra are shown by the blue and red curves in Fig. 2a for $BM^{(1)}$ and $BM^{(2)}$, respectively. As expected from effective-medium theory, $BM^{(1)}$ almost entirely governs the long-wavelength response of the grating. At $\lambda = 800$ nm, its effective index is close to 2, nearly equal to the geometric mean of the refractive indices of air and silicon, so that the grating behaves as a conventional quarter-wave antireflection coating.

The behavior at shorter wavelengths is markedly different. Remarkably, the reflectance predicted by retaining only $BM^{(2)}$ quantitatively reproduces the RCWA results, including the pronounced minimum near $\lambda = 450$ nm. This mode therefore provides the dominant channel responsible for broadband antireflection in the visible. At this wavelength, its effective index is approximately 0.7, demonstrating that efficient antireflection can be achieved with an effective index well below unity. This behavior is reminiscent of a lossless Drude plasma above its plasma frequency, where the refractive index lies between 0 and 1 and thin films may exhibit perfect antireflection through Fabry–Perot interference (Section 13.3 of Ref. [47]).

The green curve in Fig. 2a shows the reflectance obtained by retaining both propagating Bloch modes. Besides the coupling between external plane waves and Bloch modes, this two-mode model also includes the coupling between $BM^{(1)}$ and $BM^{(2)}$ at both interfaces (Fig. 2(b5)). Since these are the only propagating Bloch modes supported by the grating over the considered spectral range, the grating response is almost entirely governed by their excitation, propagation, and mutual coupling and the resulting spectrum is almost indistinguishable from the full RCWA solution (black dots).

Importantly, the S-matrix reformulation itself introduces no modal approximation. The approximation enters only when the full modal basis is deliberately truncated to a small number of propagating Bloch modes.

Although illustrated here for normal incidence, the same methodology naturally extends to oblique incidence. Additional asymmetric Bloch modes must then be included, allowing the interpretation of high-(Q) resonances and other angle-dependent phenomena reported for high-contrast gratings [29].

More generally, this example illustrates that the modal scattering coefficients provide a rigorous and intuitive framework for interpreting high-contrast gratings, which are not easily understood with coupled Mie resonances. They play the same role for periodic media as Fresnel coefficients do for homogeneous interfaces, revealing the dominant propagation channels and the coupling mechanisms responsible for the optical response.

## 4. Effective computation for multiple layer thicknesses

The strength of modal methods is their analytical treatment of propagation. In RCWA, once the Bloch modes are known, the S-matrix can be computed analytically for arbitrary layer thicknesses, as the thickness affects only the propagation phase factors. The reformulation makes this dependence directly reusable when a layer is embedded in different multilayer assemblies.

We have therefore developed a vectorized MATLAB function, **retfp.m** (where 'fp' stands for Fabry–Perot), which computes in parallel the total S-matrix of a multilayer structure for many values of the thickness of a single layer. Owing to this vectorization, the computational cost of evaluating hundreds of thicknesses is only moderately larger than that of a single-thickness calculation.

To illustrate this capability, we consider the structure shown in Fig. 3a. It consists of two silicon gratings embedded in silica and separated by a thin silica spacer. The geometry is inspired by broadband absorber designs for solar-cell applications. Each grating is formed by a periodic array of silicon Swiss crosses of different sizes. The structure is illuminated by a TM-polarized plane wave at an oblique incidence with a wavelength of $\lambda = 600$ nm.

Figure 3b presents the absorption as a function of the thickness of the upper grating, adjacent to the incident medium. The absorption is computed for $N = 400$ independent thickness values ranging from 0 to 600 nm. Figure 3c shows the computation time, normalized to that required for a single-thickness calculation, as a function of $N$. The computational overhead remains remarkably small up to about $N = 100$, staying below a factor of five. For $N = 100$ thickness values, the total computation takes only two times the single-thickness calculation. A pronounced increase appears only for $N \approx 1000$, most likely because the available RAM of the laptop becomes the limiting factor. On a workstation with a larger memory capacity, this transition is expected to occur at significantly larger values of $N$. In practice, evaluating a few hundred thicknesses is already sufficient for most optimization problems, except for extremely high-$Q$ resonant cavities.

The practical impact can be substantial. For example, a previous RCWA study of a high-Q photonic-crystal cavity required approximately $10^3$ cavity lengths and several hundred wavelengths, (see Fig. 1c of [48] and the corresponding discussion around slide 11 of [11]). The vectorized implementation reduced this otherwise prohibitive parameter sweep to a practical computation.

More generally, this illustrates a distinctive advantage of modal methods for design and optimization. In RCWA, the expensive part of the calculation—the determination of the eigenmodes—is performed only once. Geometrical parameters such as the thickness of a layer then enter analytically through the propagation operator, making it possible to evaluate hundreds or even thousands of parameter values with only a modest additional computational cost. In contrast, brute-force methods based on a full discretization of Maxwell's equations, such as the finite-difference time-domain (FDTD) or finite-element method (FEM), require a complete electromagnetic simulation for every parameter value. Consequently, parameter sweeps that are inexpensive in RCWA can become substantially more costly with these approaches. This ability to explore large parameter spaces efficiently makes modal methods particularly attractive for design, optimization, and inverse-design workflows involving repeated evaluation of layer thicknesses or other propagation-related parameters.

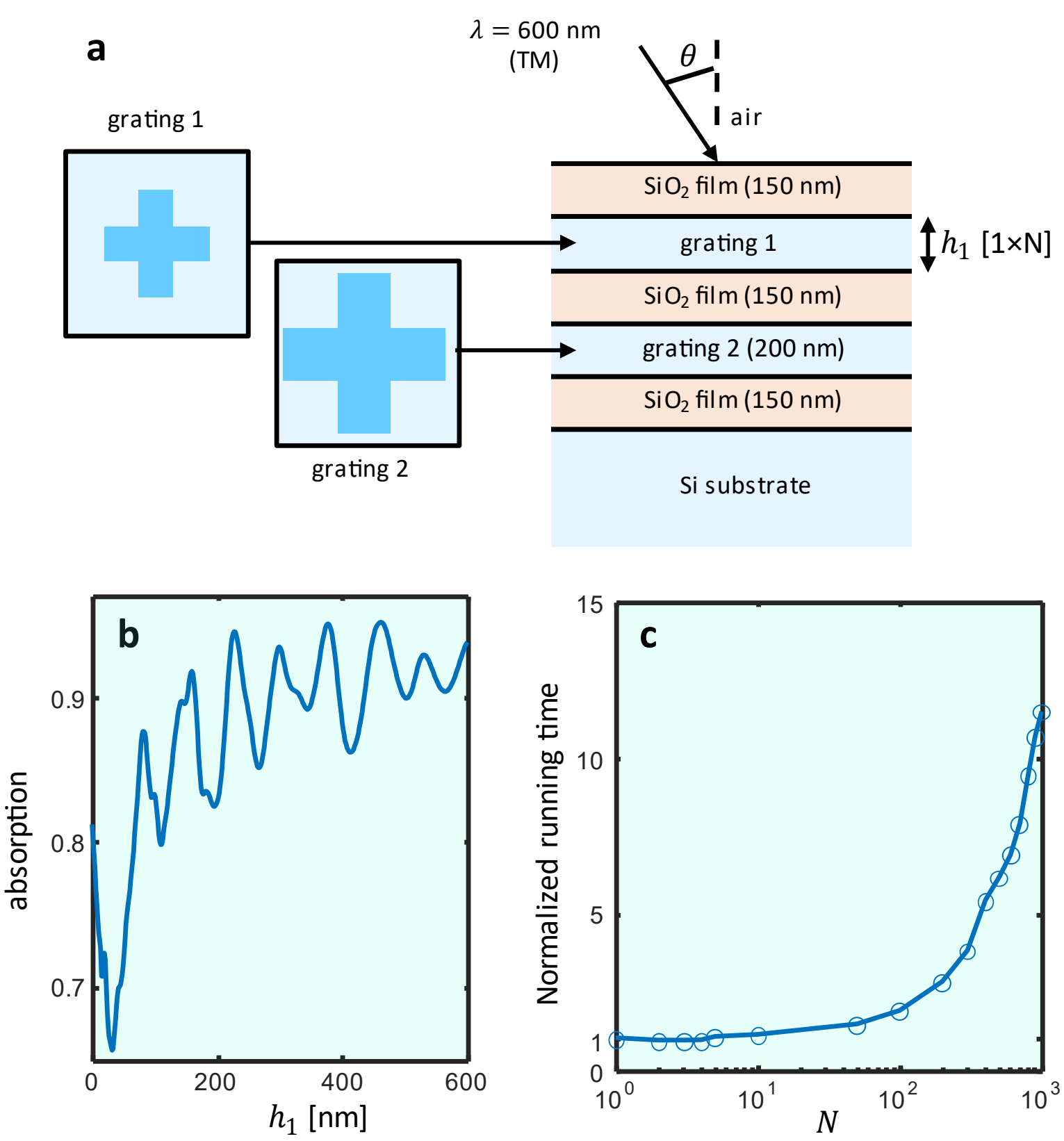

**Figure 3. Fast evaluation of thickness-dependent optical response. a** Layered grating geometry used for validation. The structure consists of two silicon gratings composed of Swiss crosses of different sizes, separated by a silica spacer. The grating periods are 1 $\mu$m in both directions. **b** Absorption in the silicon gratings and substrate (silicon refractive index $4.0506 + 0.0792i$) as a function of the thickness $h_1$ of the upper grating. **c** Computational time as a function of the number $N$ of thickness values in the vector $h_1$. The computational time is normalized to that required for a single-thickness computation. The structure is illuminated by a plane wave at $\lambda = 600$ nm at an incidence polar angle of $\theta = 60°$ with an azimuthal angle of $\delta = 20°$. Figure c can be reproduced with the documented MATLAB script **fig3retfp.m**, available upon request. Results are obtained for 31×31 retained Fourier harmonics.

Beyond design optimization, the capability to compute the response for many layer thicknesses also addresses a practical situation that frequently arises in optics. When one or several layers are much thicker than the temporal coherence length of the illumination source—as is commonly the case for the substrate in photovoltaic devices—the electromagnetic propagation is only partially coherent. Diffraction remains coherent within the gratings and thin films, whereas the multiple reflections inside the thick layer must be treated incoherently.

Incorporating this partial incoherence into an S-matrix formalism is not straightforward. A rigorous treatment requires evaluating infinite geometric series that describe the repeated incoherent coupling of the various diffracted orders between the front and rear interfaces of the thick layer [49-50].

For users who have not implemented such advanced formalisms, incoherence is often approximated by spectral averaging. A simple and often effective approximation consists of averaging the optical response over many values of the layer thickness, a strategy that is particularly attractive with the present vectorized implementation.

This approach is illustrated in Fig. 4a for a structure composed of a square array of air holes etched into a gold film deposited on a thick silica substrate whose rear surface is coated with a gold mirror. The rear coating is intentionally chosen to be highly reflective in order to demonstrate that the thickness-averaging approach remains accurate even in the presence of strong incoherent back reflections. The unit-cell dimensions are $2100 \times 2200$ nm$^2$, and the incident wavelength is 950 nm. Under these conditions, several diffraction orders propagate inside the silica substrate, making the incoherent multiple-scattering problem significantly more demanding than in the single-order case. Consequently, the diffraction efficiency is no longer described by a simple Fabry–Perot oscillation. Figure 4b shows the diffraction efficiency $\eta_{0,0}$ of the specular reflected order as a function of the substrate thickness. The response exhibits large, non-sinusoidal oscillations resulting from the interference of several propagating diffraction orders.

Figure 4c summarizes the average values of $\eta_{0,0}$ computed from 500 uniformly distributed thicknesses over six different thickness intervals. Despite the large and complex oscillations observed in Fig. 4b, the average diffraction efficiency remains nearly constant from one interval to another. This demonstrates that thickness averaging accurately reproduces the effect of temporal incoherence, even in a configuration involving multiple propagating diffraction orders and strong rear reflections. This capability offers a practical route to treating incoherent thick layers without implementing dedicated incoherent S-matrix formalisms.

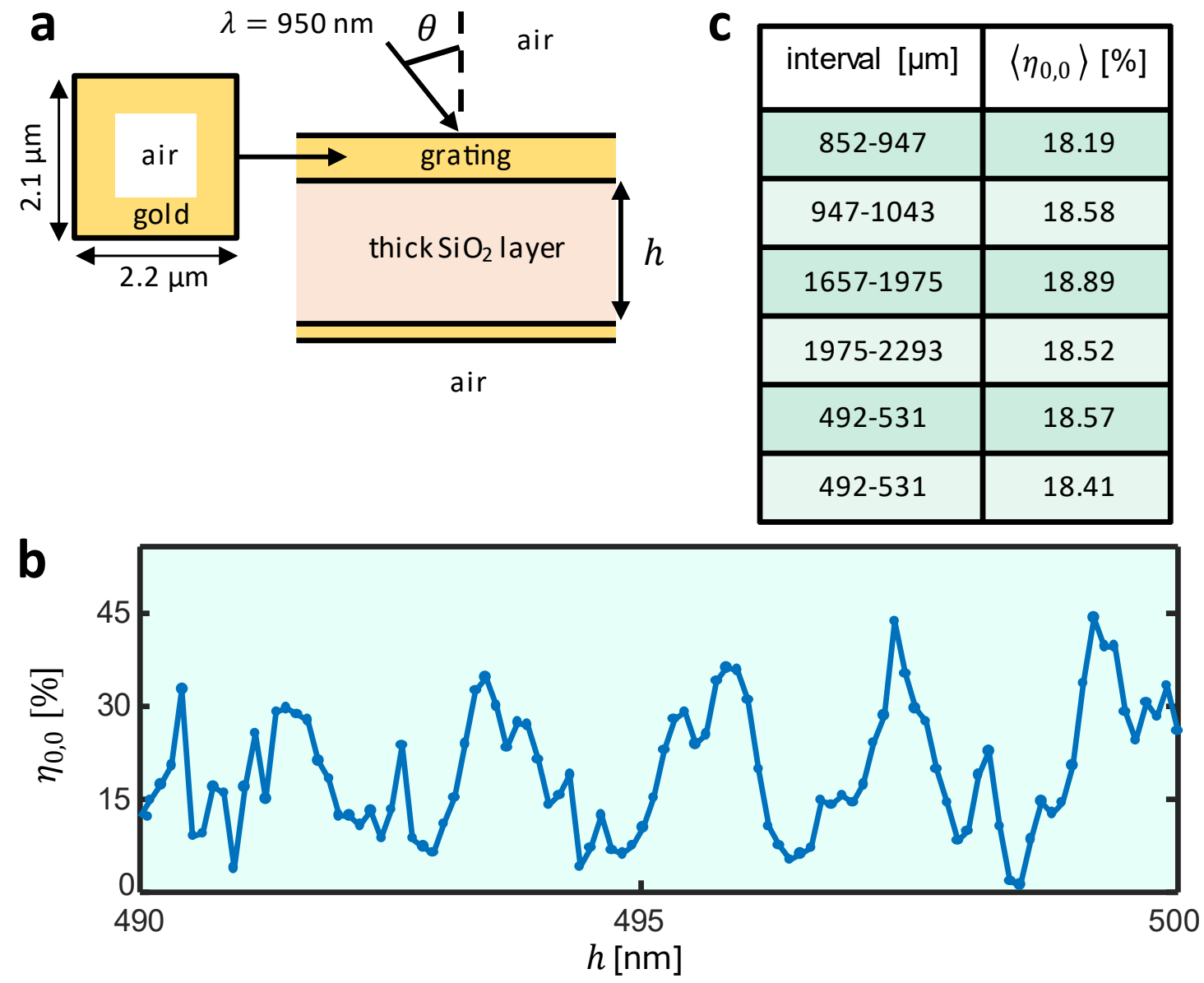

| interval [μm] | $\langle \eta_{0,0} \rangle$ [%] |
|---|---|
| 852-947 | 18.19 |
| 947-1043 | 18.58 |
| 1657-1975 | 18.89 |
| 1975-2293 | 18.52 |
| 492-531 | 18.57 |
| 492-531 | 18.41 |

**Figure 4. Temporal incoherence modeled by thickness averaging. a** Geometry of the layered grating. A periodic array of rectangular air holes is etched into a gold layer. The grating periods are $a_x = 2.2$ μm and $a_y = 2.1$ μm. The air holes have dimensions $0.7a_x$ along the $x$-direction and $0.7a_y$ along the $y$-direction. The structure is deposited on a thick silica substrate whose rear interface is coated with a 70 nm-thick gold film. The gold refractive index at $\lambda = 950$ nm is $0.235 + 6.438i$. The structure is illuminated by a TM polarized plane wave at an incidence polar angle of $\theta = 60°$with an azimuthal angle of $\delta = 20°$. **b** Specular diffraction efficiency, $\eta_{0,0}$, as a function of the silica substrate thickness $h$ over a 10 nm interval (490–500 nm). The response exhibits strong, non-sinusoidal oscillations because several diffraction orders propagate within the substrate and interfere after reflection at the rear gold mirror. **c** Average values of $\eta_{0,0}$ computed from 500 uniformly distributed thicknesses exhibit only a weak dependence on the thickness windows. The figure can be reproduced with the documented MATLAB script **fig4retfp.m**, available upon request.

## 5. S-matrix computation for multiple incidences

The intrinsic S-matrix formulation also provides a convenient representation in which several incident plane-wave channels can be treated simultaneously. Conventional RCWA implementations [10,53] and most available software packages [13] compute the diffraction efficiencies associated with a single incident plane wave. In contrast, the S-matrix naturally relates all incident and outgoing plane-wave ports. Once the eigenproblem has been solved for a given in-plane Bloch wavevector $\mathbf{k}_\parallel$, the scattering coefficients between all propagating plane waves with wavevectors $\mathbf{k}_\parallel + (mK_x\hat{\mathbf{x}} + nK_y\hat{\mathbf{y}})$ are obtained simultaneously.

This property becomes particularly advantageous when the grating period is comparable to or larger than the wavelength, so that several diffraction orders are propagating. In this case, a single RCWA computation provides the optical response for many incident directions at once, these directions being related by the grating equations. The same situation naturally arises in supercell simulations of disordered metasurfaces, where large artificial periods are introduced to preserve randomness while applying periodic boundary conditions [14]. Applications include broadband antireflection coatings for solar cells [54], bidirectional reflectance distribution function (BRDF) calculations for appearance engineering [55], radiative cooling [56], and the homogenization of disordered metamaterials [57].

We have implemented this multi-incidence capability in RETICOLO. It requires only a reformulation of the plane-wave basis and leaves the S-matrix machinery unchanged: the modal computation, the Redheffer star-product algorithm, and the numerical stability are identical to those of the conventional implementation. The only additional cost is the manipulation of larger scattering matrices, which remains modest compared with the computation of the eigenmodes. This capability is a direct consequence of the modal formulation, in which propagation through each layer is treated analytically, and is particularly natural when the complete set of scattering channels is available after solving the eigenproblem.

Figure 5 illustrates the method for a grating composed of square air holes etched into a semiconductor layer (Fig. 5a), with periods $a_x = 2.1\lambda$ and $a_y = 2.2\lambda$. Two computations are compared.

In the first approach, we use the multi-incidence scheme, with $\mathbf{k}_{inc} = [k_x, k_y] = n_h\omega/c\ \sin\theta\ [\cos\delta, \sin\delta]$, where $n_h = 1$ is the refractive index of air. We compute the diffraction efficiency of the TE-polarized plane wave propagating in diffraction order $(m = -1, n = -1)$ when the grating is illuminated by a TM polarized plane wave with an in-plane wavevector $[k_x + K_x, k_y - K_y]$. The resulting efficiency is plotted as circles in Fig. 5b as a function of the truncation rank $M$, corresponding to a Fourier expansion with $2M + 1$ coefficients along both the $x$ and $y$ directions.

For comparison, a conventional RCWA computation is performed with an incident wavevector $\mathbf{k}_{inc} = n_h k_0 \sin\theta\, [\sin\delta, \cos\delta] + [2\pi/a_x, -2\pi/a_y]$. The corresponding diffraction efficiency is shown by the blue curve. Although formulated in the standard framework, this calculation describes exactly the same physical scattering process: the in-plane wavevectors of both the incident and diffracted plane waves are identical to those of the multi-incidence computation, but they are now identified as the incident order $(0,0)$ and the diffracted order $(-2,0)$ in the air superstrate. As shown in Fig. 5b, the two approaches are numerically indistinguishable, thereby validating the proposed implementation.

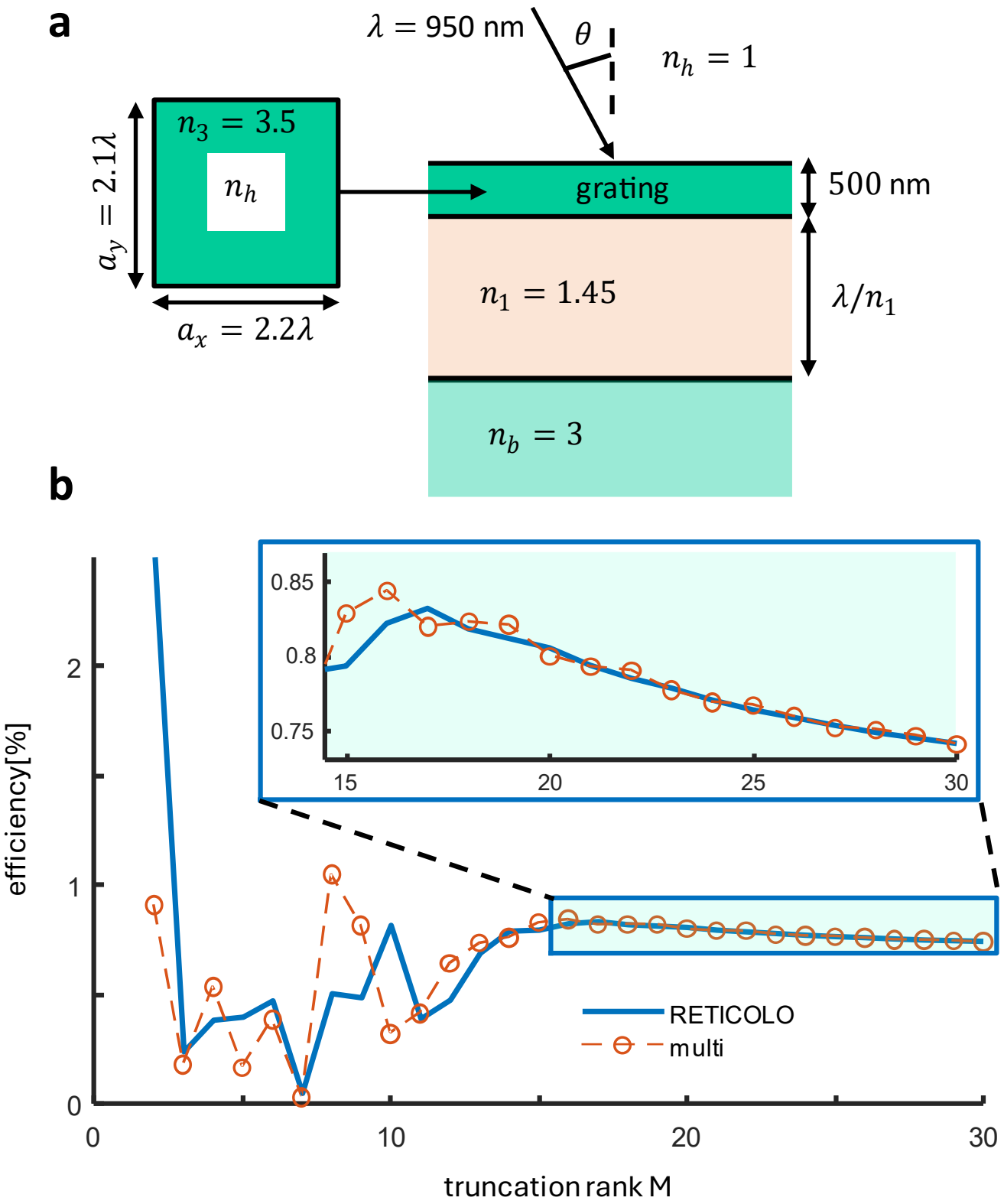

**Figure 5. S-matrix computation with multi-incident plane waves. a** Grating geometry used for validation. The rectangular air hole has dimensions $0.7a_x$ along $x$ and $0.7a_y$ along $y$. **b** Convergence versus truncation rank $M$. The circles correspond to calculations performed for $\mathbf{k}_{inc} = n_h k_0 \sin\theta\,[\sin\delta\,,\cos\delta]$, $k_0 = 2\pi/\lambda$. They represent the transmission efficiency into a TE-polarized plane wave in the air superstrate with in-plane wavevector $\mathbf{k}_{inc} - [2\pi/a_x, 2\pi/a_y]$, i.e. diffraction order $(-1,-1)$, for a TM-polarized incident plane wave with in-plane wavevector $\mathbf{k}_{inc} + [2\pi/a_x, -2\pi/a_y]$, i.e. order $(1,-1)$, illuminating the grating from the substrate. The blue line corresponds to calculations performed for $\mathbf{k}_{inc} = n_h k_0 \sin\theta\,[\sin\delta\,,\cos\delta] + [2\pi/a_x, -2\pi/a_y]$. The in-plane wavevectors are identical to those of the previous case, but now correspond to the incident order (0,0) and the diffracted order $(-2,0)$ in the air superstrate (classical RCWA formulation). Here, $\theta = 60°$, $\delta = 60°$ and $\lambda = 950$ nm. The figure can be reproduced using the documented MATLAB function **fig5multi_illumination.m**, available upon request, together with the companion script **run_fig5multi_illumination.m**.

## 6. Conclusion

The combination of RCWA with the S-matrix formalism has enjoyed more than three decades of well-deserved success for the analysis and design of diffraction gratings and metasurfaces. In this work, we have introduced a slight reformulation of the S-matrix algorithm that preserves its numerical stability and computational efficiency while redefining the S-matrix as an intrinsic scattering operator associated with a single layer rather than with two adjacent layers. This separation of intrinsic layer propagation from interface-dependent coupling is the central feature of the reformulation: each layer can be characterized independently and subsequently reused in complicated multilayer assemblies [19], without recomputing interface-dependent quantities.

The reformulation provides direct access to the coupling coefficients governing energy exchange between plane waves and Bloch modes at periodic interfaces. These generalized Fresnel coefficients rigorously incorporate the electromagnetic boundary conditions and provide physical insight into wave propagation in periodic layered systems and facilitate their design, as illustrated with the tutorial example of a broadband antireflection grating.

Beyond this conceptual simplification, the reformulation also brings significant computational benefits. It enables the simultaneous evaluation of the optical response for many layer thicknesses at nearly the cost of a single simulation, thereby greatly accelerating parameter sweeps and optimization. It also allows the grating response to be computed for multiple incident wavevectors simultaneously, an important advantage for structures with large periods or for supercell-based designs.

These new capabilities have been implemented in the RETICOLO freeware. The documented MATLAB scripts used to generate all figures are currently available from the authors upon request. Following publication, they will be incorporated into a new public release of RETICOLO.

**Annex 1: Implementation within RETICOLO environment**

Table 1 summarizes the main scripts provided by the freeware for S-matrix manipulations.

**Table 1.** Main RETICOLO functions used to implement S-matrix algebra.

| MATLAB script | description |
|---|---|
| **retc.m** | $\mathrm{S} = \mathrm{retc}(a, h)$<br>computes the S-matrix of a layer characterized by a thickness $h$ and by a descriptor $a$. |
| **retss.m** | $\mathrm{S} = \mathrm{retss}(s_1, s_2, s_3)$<br>computes the S-matrix obtained from the cascade Redheffer product $\mathrm{S} = s_1 \star s_2 \star s_3$. |
| **retb.m** | $\mathrm{S} = \mathrm{retb}(init, a, sens)$<br>computes the change-of-basis operator that transforms field coefficients from the Fourier basis to the modal basis for the layer defined by the descriptor $a$. The parameter sens is a real-valued scalar whose sign determines the direction of the transformation: a positive value corresponds to the $z > 0$ direction (top interface, sh), while a negative value corresponds to the $z < 0$ direction (bottom interface, sb). |
| **retfp.m** | $\mathrm{S} = \mathrm{refp}(thickness, a, sfph, sfpb)$<br>computes the scattering matrix of a layer defined by the descriptor $a$ for multiple values of the layer thickness. The computation is performed in the modal basis of the layer by means of the change-of-basis matrices $sfph$ and $sfpb$. |
| **rettneff.m** | displays a complex-plane plot of the effective indices of all Bloch modes, with Im $(n_{\mathrm{eff}})$plotted versus Re $(n_{\mathrm{eff}})$. The user can then interactively select a subset of modes, which are subsequently plotted for further inspection. |
| **retneff.m** | sorts and selects the Bloch modes according to their effective indices. |

The script **retc.m** computes the scattering matrix of a layer characterized by its thickness and descriptor. The descriptor is a MATLAB structure containing the modal fields and propagation constants of the layer, obtained from the eigenvalue diagonalization.

The script **retss.m** performs S-matrix cascades using the Redheffer star product [12,4], commonly denoted by $\star$ to distinguish it from the conventional matrix product. The star product is applied sequentially, starting from the layer adjacent to the superstrate and ending with the layer adjacent to the substrate. Since it is associative, multilayer structures can be assembled recursively in any convenient order. Furthermore, when the elementary scattering matrices satisfy reciprocity or energy-conservation relations, these properties are preserved by the resulting global scattering matrix. Consequently, the Redheffer star product provides not only a numerically stable algorithm for cascading layers, but also a rigorous and physically transparent composition law for modal scattering processes.

All these operations are initially performed in the Fourier basis. The script **retb.m** provides the change-of-basis operators required to transform scattering matrices from the Fourier basis to the modal basis. In the companion codes, **retb.m** generates two change-of-basis matrices, denoted $sh$ and $sb$, depending on whether the transformation is performed toward the modes of the top layer ($sh$) or the bottom layer ($sb$). The final mode-to-mode scattering matrix is then obtained as $S = \mathrm{retss}(sh, s_1, s_2, s_3, sb)$, where $s_p$denotes the scattering matrix of the $p$-th layer.

To further assist users, RETICOLO version 11 and later are distributed together with this report and include all the scripts used to generate the figures presented in this article. These scripts are fully documented and are intended to serve as practical examples of the software capabilities, while also providing a useful starting point for users wishing to adapt the code to their own applications.

**Acknowledgements**. The authors acknowledge fruitful interactions with many colleagues at Institut d’Optique and Stanford University, especially Nayeun Lee (now at Apple), Anqi Ji (now at Meta), Qitong Li (now at Tsinghua University) and Mark Brongersma.

**Disclosures**. The authors declare no conflicts of interest.

**References**

1. T. Itoh, "Numerical Techniques for Microwave and Millimeter-wave Passive Structures", John Wiley & Sons, New York (1989).
2. N. Château and J.-P. Hugonin, "Algorithm for the rigorous coupled wave analysis of grating diffraction", J. Opt. Soc. Am. A **11**, 1321-1331 (1994).
3. N. P. K. Cotter, T. W. Preist and J. R. Sambles, "Scattering-matrix approach to multilayer diffraction", J. Opt. Soc. Am. A **12**, 1097-1103 (1995).
4. L. Li, "Formulation and comparison of two recursive matrix algorithms for modeling layered diffraction gratings," J. Opt. Soc. Am. A **13**, 1024–1035 (1996).
5. E. L. Tan, "Note on formulation of the enhanced scattering- (transmittance-) matrix approach," J. Opt. Soc. Am. A **19**, 1157-1161 (2002).
6. R. Pregla and W. Pasher, "The Method of Lines", in *Numerical Techniques for Microwave and Millimeter Wave Passive Structures*, T. Itoh, ed. (Wiley, New York, 1989), pp. 381-446.
7. J.-P. Hugonin, M. Besbes, P. Lalanne, "Hybridization of electromagnetic numerical methods through the G-matrix algorithm", Opt. Lett. **33**, 1590-1592 (2008).
8. J.-P. Hugonin and P. Lalanne, RETICOLO software for grating analysis, Institut d'Optique, Orsay, France (2005), arXiv:2101.00901v4.
9. P. Lalanne and G. M. Morris, "Highly improved convergence of the coupled-wave method for TM polarization", J. Opt. Soc. Am. A **13**, 779-784 (1996).
10. L. Li, "New formulation of the Fourier modal method for crossed surface-relief gratings", J. Opt. Soc. Am. A **14**, 2758-2767 (1997).
11. P. Lalanne, webinar on freeware organized by GDR Ondes, "Some electromagnetic freeware at Institut d’Optique", February 5 (2025) https://www.youtube.com/watch?v=2f6B9Ka3mJE
12. R. Redheffer, "Difference equations and functional equations in transmission-line theory", in *Modern Mathematics for the Engineer,* E. F. Beckenbach, ed. (McGraw-Hill, New York, 1961), Chap. 12, pp. 282–337.

13. V. Liu and S. Fan, "S$^4$: A free electromagnetic solver for layered periodic structures", Computer Physics Communications **183**, 2233-2244 (2012).
14. P. Lalanne, M. Chen, C. Rockstuhl, A. Sprafke, A. Dmitriev, K. Vynck, "Disordered optical metasurfaces: basics, properties, and applications", Adv. Opt. Photon. **17**, 45-112 (2025).
15. I. M. Fradkin, S. A. Dyakov and N. A. Gippius, "Nanoparticle lattices with bases: Fourier modal method and dipole approximation", Phys. Rev. B **102**, 045432 (2020).
16. G. Lecamp, J.-P. Hugonin and P. Lalanne, "Theoretical and computational concepts for periodic optical waveguides", Opt. Express **15**, 11042-60 (2007).
17. E. Silberstein, P. Lalanne, J.-P. Hugonin and Q. Cao, "Use of grating theory in integrated optics", J. Opt. Soc. Am. A. **18**, 2865-75 (2001).
18. G. Lecamp, P. Lalanne and J.-P. Hugonin, "Very large spontaneous emission beta-factors in photonic crystal waveguides", Phys. Rev. Lett. **99**, 023902 (2007).
19. R. Faggiani, A. Baron, X. Zang, L. Lalouat, S.A. Schulz, Bryan O'Regan, K. Vynck, B. Cluzel, F. de Fornel, T. F. Krauss and P. Lalanne, "Lower bounds for the spatial extent of optical localized modes in 1D disordered periodic media", Sci. Rep. **6**, 27037 (2016).
20. M. G. Moharam, D. A. Pommet and E. B. Grann, T. K. Gaylord, "Stable implementation of the rigorous coupled-wave analysis for surface-relief gratings: enhanced transmittance matrix approach", J. Opt. Soc. Am. A **12**, 1077-1086 (1995).
21. D. M. Whittaker and I. S. Culshaw, "Scattering-matrix treatment of patterned multilayer photonic structures", Phys. Rev. B **60**, 2610-2618 (1999).
22. A. Gras, W. Yan, P. Lalanne, "Quasinormal-mode analysis of grating spectra at fixed incidence angles", Opt. Lett. **44**, 3494-3497 (2019).
23. S. Kim, Y. G. Peng, S. Yves, A. Alù, "Loss Compensation and Superresolution in Metamaterials with Excitations at Complex Frequencies", Phys. Rev. X **13**, 041024 (2023).
24. F. Guan, N. Chen, Z. Lin, W. Song, S. Zhu, T. Li, S. Zhang, "Spectral window engineering for synthetic wave compensation of plasmonic loss", arXiv:2604.26628
25. P. Lalanne, T. Wu, "Complex-frequency superlensing faces intrinsic limitations", Optica **13**, 1377-1386 (2026).
26. H. Benisty, J. J. Greffet and P. Lalanne, *Introduction to Nanophotonics* (Oxford University Press, Oxford, 2022).
27. T. Clausnitzer, T. Kämpfe, E.-B. Kley, A. Tünnermann, U. Peschel, A.V. Tishchenko, and O. Parriaux, "An intelligible explanation of highly-efficient diffraction in deep dielectric rectangular transmission gratings", Opt. Express **13**, 10448-10456 (2005)
28. A. V. Tishchenko, "Phenomenological representation of deep and high contrast lamellar gratings by means of the modal method", Opt. Quantum Electron. **37**, 309-330 (2005)
29. P. Lalanne, J.-P. Hugonin, P. Chavel, "Optical properties of deep lamellar gratings: a coupled Bloch-mode insight", IEEE J. Lightwave Technol. **24**, 2442-2449 (2006)
30. V. Karagodsky and C. J. Chang-Hasnain, "Physics of near-wavelength high contrast gratings," Opt. Express **20**, 10888-10895 (2012)
31. A. I. Ovcharenko, C. Blanchar, J.-P. Hugonin and C. Sauvan, "Bound states in the continuum in symmetric and asymmetric photonic crystal slabs", Phys. Rev. B **101**, 155303 (2020).
32. D. A. Bykov, E. A. Bezus, L. L. Doskolovich, "Coupled-wave formalism for bound states in the continuum in guided-mode resonant gratings", Phys. Rev. A **99**, 063805 (2019).
33. K. L. Koshelev, Z. F. Sadrieva, A. A. Shcherbakov, Y. S. Kivshar, A. A. Bogdanov, "Bound states in the continuum in photonic structures", Phys. Usp. **66**, 494-517 (2023).
34. O. Parriaux, N. M. Lyndin, "Modal phenomenology of arbitrarily narrow reflection from high contrast binary gratings", J. Opt. **21**, 085608 (2019).
35. J. Liu, Z. Peng, Q. Song, A. Chen, L. Yang, C. Zheng and D. Han, "Complex band structure and bound states in the continuum: a unified theoretical framework", Rep. Prog. Phys. **89** 037901 (2026).

36. J. H. Song, P. Lalanne, M. K. Seo, M. L. Brongersma, "Transfer Matrix Method-Compatible Model for Metamaterial Stacks", ACS Photonics **10**, 2948-2954 (2023)
37. M. A. Green and M. J. Keevers, "Optical properties of intrinsic silicon at 300 K", Prog. Photovolt: Res. Appl. **3**: 189-192 (1995).
38. N. Lee, M. Xue, J. Hong, J. van de Groep, M. L. Brongersma, Multi-Resonant Mie Resonator Arrays for Broadband Light Trapping in Ultrathin c-Si Solar Cells. Adv. Mater. **35**, 2210941 (2023).
39. W. Liu and Y. S. Kivshar, "Generalized Kerker effects in nanophotonics and meta-optics", Opt. Express **26**, 13085-13105 (2018).
40. P. Spinelli, M. A. Verschuuren, A. Polman, "Broadband omnidirectional antireflection coating based on subwavelength surface Mie resonators", Nat. Commun. **3**, 692 (2012).
41. A. Cordaro, J. Van De Groep, S. Raza, E. F. Pecora, F. Priolo, M. L. Brongersma, "Antireflection High-Index Metasurfaces Combining Mie and Fabry-Pérot Resonances", ACS Photonics **6**, 453 (2019).
42. S. Collin, "Nanostructure arrays in free-space: Optical properties and applications", Rep. Prog. Phys. **77**, 126402 (2014).
43. I. Karavaev, K. Koshelev, and A. Bogdanov, "Optical Resonances: From Eigenmodes to Scattering Features", arXiv:2603.13845v1 (2026)
44. C. F. Bohren and D. R. Huffman, *Absorption and scattering of light by small particles*, (Wiley 1983).
45. C. Gigli, Q. Li, P. Chavel, G. Leo, M. Brongersma, P. Lalanne, "Fundamental limitations of Huygens' metasurfaces for optical beam shaping", Laser Photonics Reviews **15**, 2000448 (2021).
46. R. Dezert, P. Richetti, A. Baron, "Complete multipolar description of reflection and transmission across a metasurface for perfect absorption of light", Opt. Express **27**, 26317-26330 (2019).
47. M. Born and E. Wolf, "Principle of Optics", 6th ed. (Macmillan, New York, 1964).
48. P. Lalanne, S. Mias, J.-P. Hugonin, "Two physical mechanisms for boosting the quality factor to cavity volume ratio of photonic crystal microcavities", Opt. Express **12**, 458-467 (2004).
49. N. Chateau, J.-P. Hugonin, B. Guldimann, P. Chavel, "Two-wave diffraction of quasi-monochromatic light by a volume grating deposited on a thick transparent plate", Opt. Comm. **103**, 444-452 (1993).
50. K. Postava, T. Fordos, T. Kohut and L. Halagačka, "Coherent and Incoherent Phenomena in Anisotropic Periodic Gratings," *2025 International Conference on Numerical Simulation of Optoelectronic Devices (NUSOD)*, Lodz, Poland, pp. 57-58 (2025), doi: 10.1109/NUSOD64393.2025.11199617
51. M. G. Moharam and T. K. Gaylord, "Rigorous coupled-wave analysis of grating diffraction - E-mode polarization and losses", J. Opt. Soc. Am. **73**, 451-455 (1983).
52. R. H. Morf, "Exponentially convergent and numerically efficient solution of Maxwell's equations for lamellar gratings", J. Opt. Soc. Am. A **12**, 1043-1056 (1995).
53. M. G. Moharam, E. B. Grann, D. A. Pommet and T. K. Gaylord, "Formulation for stable and efficient implementation of the rigorous coupled-wave analysis of binary gratings", J. Opt. Soc. Am. A **12**, 1068-1076 (1995).
54. N. Tucher, H.T. Gebrewold and B. Bläsi, "Field stitching approach for the wave optical modeling of black silicon structures", Opt. Express **26**, 332840 (2018).
55. K. Vynck, R. Pacanowski, A. Dufay, X. Granier, and P. Lalanne, "The visual appearances of disordered metasurfaces", Nature Materials **21**, 1035-41 (2022).
56. R. Liu, S. Wang, Z. Zhou, K. Zhang, G. Wang, C. Chen, Y. Long, "Materials in Radiative Cooling Technologies", Adv. Mater. 37, 2401577 (2025).

57. D. Theobald, D. Beutel, L. Borgmann, H. Mescher, G. Gomard, C. Rockstuhl, U. Lemmer, "Simulation of light scattering in large, disordered nanostructures using a periodic T-matrix method", J. Quant. Spectrosc. Radiat. Transf. **272**, 107802 (2021).